\documentclass[preprint,superscriptaddress,longbibliography]{revtex4-1}
\usepackage{graphicx}
\usepackage{epstopdf, epsfig}
\usepackage{hyperref}

\usepackage{amssymb}
\usepackage{subfigure}
\usepackage{amsmath}

\usepackage{graphicx}
\usepackage{epstopdf}
\usepackage{color}
\usepackage{ulem}

\begin{document}

\title{Analytical Calculation of the Orbital Spectrum of the Guiding Center Motion in Axisymmetric Magnetic Fields}

\author{Yannis Antonenas}
\affiliation{School of Applied Mathematical and Physical Sciences, National Technical University of Athens, Athens, Greece}

\author{Giorgos Anastassiou}
\affiliation{School of Electrical and Computer Engineering, National Technical University of Athens, Athens, Greece}

\author{Yannis Kominis}
\affiliation{School of Applied Mathematical and Physical Sciences, National Technical University of Athens, Athens, Greece}


\begin{abstract}
Charged particle motion in axisymmetric toroidal magnetic fields is analyzed within the context of the canonical Hamiltonian Guiding Center theory. A canonical transformation to variables measuring the drift orbit deviation from a magnetic field line is introduced and an analytical transformation to Action-Angle variables is obtained, under a zero drift width approximation. The latter is used to provide compact formulas for the orbital spectrum of the drift motion, namely the bounce/transit frequencies as well as the bounce/transit averaged toroidal precession and gyration frequencies. These formulas are shown to have a remarkable agreement with numerically calculated full drift width frequencies and significant differences with standard analytical formulas based on a pendulum-like Hamiltonian description. The analytical knowledge of the orbital spectrum is crucial for the formulation of particle resonance conditions with symmetry breaking perturbations and the study of the resulting particle, energy and momentum transport. 
\end{abstract}

\maketitle

\section{Introduction}
Charged particle dynamics in toroidal magnetic fields has been the key theoretical issue for the study of magnetically confined fusion plasmas for many decades. The understanding of the role of the magnetic field topology on single as well as on collective particle dynamics has been crucial for the design of fusion devices with magnetic fields having different types of symmetries \citep{Freidberg}. These background magnetic fields also determine the interactions between particles and symmetry-breaking perturbations of a large range of spatial and temporal scales, mainly through resonance conditions, and the corresponding particle, energy and momentum transport.

The Guiding Center (GC) theory \citep{Littlejohn_83} has been used for a rigorous dynamical reduction by the systematic elimination of the rapidly varying gyro-angle variable related to the cyclotron motion around a magnetic field line. In the GC description, the magnetic moment is a constant of the motion and the particle dynamics are studied in terms of the drift motion of the center of the cyclotronic motion, leading to a gyro-kinetic Hamiltonian theory. Although the original derivation of the GC equations has been formulated in non-canonical variables, the utilization of magnetic coordinates has been shown to allow a Hamiltonian formulation in canonical variables \citep{White_84, Littlejohn_85, White_book}. 

The Hamiltonian formulation of the guiding center motion in canonical variables has several computational and conceptual advantages that are  revealed under a transformation to Action-Angle variables \citep{Goldstein, Li-Li}. Such a transformation is possible for the case of GC motion in an axisymmetric magnetic field where the Hamiltonian system is integrable. In the Action-Angle variable set, the topology of the motion is described by multi-dimensional tori and each orbit is labeled by a distinct set of the invariant values of the three Action variables \citep{Kaufman_72}. This simple orbit parametrization allows for an orbit-based analysis of particle, energy and momentum transport which is particularly useful for the study of energetic particle dynamics in fusion plasmas in direct relation to velocity-space tomography techniques \citep{Stagner_17, Tholerus_17}. Moreover, all the orbital frequencies can be readily calculated in terms of the Action variables. The latter determine the resonance conditions between particles and symmetry-breaking perturbations \citep{Zestanakis_16, Bierwage_16} resulting in breaking of the Action invariance and diffusion in the Action space, that describes energy, momentum and radial particle transport \citep{Kominis_08a, Kominis_08b, Kominis_10, Petrov_16}. The collective particle dynamics, under the presence of perturbations is characterized by the modification of the unperturbed particle distribution functions \citep{White_12, White_11, Podesta_14}. Another important feature of the Action-Angle description is that the different time scales of the motion are well separated in different degrees of freedom allowing for a systematic dynamical reduction to a hierarchy of evolution equations for the reduced distribution functions \citep{Brizard_00, Kominis_10}. 

In this work, we utilize the canonical Hamiltonian formulation of the GC motion along with an appropriate canonical transformation in order to facilitate the calculation of Action-Angle variables for the case of an axisymmetric magnetic field. By doing so we fully exploit the advantages of the canonical Hamiltonian formulation in terms of the calculation of the orbital frequencies of all degrees of freedom corresponding to bounce/transit, bounce-averaged toroidal precession and gyration frequencies, as well as the calculation of the Action variables allowing for the dynamical reduction to a bounce-averaged system. Analytical results are obtained under a Zero Drift Width (ZDW) approximation providing compact formulas for the orbital frequencies that are in remarkable agreement with Full Drift Width (FDW) numerical calculations and have significant qualitative and quantitative differences with standard analytical formulas based on a pendulum-like GC Hamiltonian.

In Section 2, the canonical formulation of the GC theory is briefly presented for completeness. In Section 3, we introduce a canonical transformation to variables measuring Drift Orbit Deviation from a given field line. The transformation is general and applies to both axisymmetric and non-axisymmetric equilibria of arbitrary shape. We show that, for a Large Aspect Ratio (LAR) magnetic field equilibrium, the transformation takes a particularly simple form and we briefly comment on possibilities of treating equilibria with higher-order terms with respect to the inverse aspect ratio. The general form of the transformation to Action-Angle variables as well as the corresponding calculation of the orbital frequencies is also presented. In Section 4, we apply a ZDW approximation for the LAR equilibrium, according to which the GC orbit is considered to take place on a single flux surface, and we obtain simple analytical formulas for the Orbital Spectrum of the GC motion. We present a novel ZDW Hamiltonian retaining terms that are significant for particles with smaller pitch angles in comparison to standard pendulum-like Hamiltonians describing deeply trapped particles, and we compare the analytical results to FDW numerical calculations. The summary and conclusions are given in Section 5.

\section{Canonical Guiding Center Hamiltonian for an Axisymmetric Equilibrium}
A general axisymmetric toroidal magnetic configuration consisting of nested toroidal flux surfaces can be represented in White-Boozer \citep{White_book} coordinates as
\begin{equation}
\mathbf{B}=g(\psi) \nabla\zeta+I(\psi) \nabla \theta+\delta(\psi,\theta) \nabla \psi_p \label{Boozer}
\end{equation}
where $\zeta$ and $\theta$ are the toroidal and the poloidal angles. The toroidal flux $\psi$ is related to the poloidal flux $\psi_p$ through the safety factor $q(\psi)=d\psi / d\psi_p$.  The functions $g$ and $I$ are related to the poloidal and toroidal currents and $\delta$ is related to the nonorthogonality of the coordinate system. 

The guiding center motion of a charged particle is described by the Lagrangian \citep{Littlejohn_83} $L=\left( \mathbf{A}+\rho_\parallel  \mathbf{B} \right) \cdot \mathbf{v} +\mu \dot{\xi}-H$, where $\mathbf{A}$ and $\mathbf{B}$ are the vector potential and the magnetic field, $\mathbf{v}$ is the guiding center velocity, $\mu$ is the magnetic moment, $\xi$ is the gyrophase, $\rho_\parallel$ is the velocity component parallel to the magnetic field, normalized to $B$, and
\begin{equation}
 H=\rho_\parallel^2B^2/2+\mu B     \label{H1}
\end{equation}
is the Hamiltonian. The guiding center motion is given in normalized units where time is normalized to $\omega_0^{-1}$, with $\omega_0=eB_0/m$ being the on-axis gyrofrequency, and distance is normalized to the major radius $R$, so that energy is normalized to $m\omega_0^2R^2$. According to the ordering of the guiding center approximation, the gyroradius is $\rho=v/B<<1$ and the magnetic moment $\mu=v_\perp^2/(2B)$ as well as the cross field drift are of order $\rho^2$ \citep{White_book}. 

The three couples of canonically conjugate variables for this GC Hamiltonian are $(\mu, \xi)$, $(P_\theta, \theta)$ and $(P_\zeta,\zeta)$  with 
\begin{eqnarray}
 P_\theta & = & \psi+\rho_\parallel I(\psi) \nonumber \\
 P_\zeta & = & \rho_\parallel g(\psi)-\psi_p(\psi). \label{canonical}
 \end{eqnarray}
providing the relation between the canonical momenta $(P_\theta, P_\zeta)$ and $(\psi,\rho_\parallel)$ \citep{White_book}. In terms of these canonical variables, Eq. (\ref{H1}) can be written as
\begin{equation}
 H(P_\theta,\theta, P_\zeta,\zeta, \mu, \xi)=\frac{\left[P_\zeta+ \psi_p(P_\theta,P_\zeta)\right]^2}{2g^2(\psi(P_\theta,P_\zeta))}B^2(\psi(P_\theta,P_\zeta),\theta)+\mu B(\psi(P_\theta,P_\zeta),\theta). \label{H2}
\end{equation}
The gyroangle $\xi$ does not appear in the guiding center Hamiltonian which is also independent of $\zeta$ due to axisymmetry of the magnetic field; therefore the corresponding canonical momenta, namely $\mu$ and $P_\zeta$, are constants of the motion and since the Hamiltonian does not depend explicitely on time (autonomous system), the system is integrable.

\section{Canonical Transformation to Drift Orbit Deviation Variables}
The guiding center Hamiltonian describes all particle drifts due to  the inhomogeneity of the magnetic field, causing the guiding center deviation from a field line. A canonical transformation with generating function \citep{Goldstein}
\begin{equation}
 \bar{F}(P_\theta, \bar{\theta}, P_\zeta,\bar{\zeta}, \mu, \bar{\xi})=-\bar{\theta} (P_\theta-P_{\theta 0})-\bar{\zeta}\left(P_\zeta+\int^{P_\theta}\frac{dP'_\theta}{q\left(\psi(P'_\theta, P_\zeta)\right)} \right) - \bar{\xi} \mu
\end{equation}
transforms to a new (barred) variable set, related to the original variables as 
\begin{eqnarray}
 \bar{P_\theta}&=&P_\theta-P_{\theta 0} \nonumber\\
 \bar{\theta}&=&\theta-\frac{\bar{\zeta}}{q(\psi(P_\theta,P_\zeta))} \nonumber \\
 \bar{P_\zeta}&=&P_\zeta+\int^{P_\theta} \frac{dP'_\theta}{q\left(\psi(P'_\theta, P_\zeta)\right)} \\
 \zeta &=&\bar{\zeta} \left( 1+\int^{P_\theta}\frac{\partial q^{-1}}{\partial \psi}\frac{\partial \psi(P'_\theta, P_\zeta)}{\partial P_\zeta} dP'_\theta \right) \nonumber \\
 \bar{\mu}&=&\mu \nonumber \\
 \bar{\xi}&=&\xi\nonumber
\end{eqnarray}

This canonical tranformation is general and can be applied for either axisymmetric or non-axisymmetric equilibrium magnetic fields. The physical meaning of the new canonical variables becomes obvious for a Large Aspect Ratio (LAR) cylindrical equilibrium described by $g=1$, $I=0$, $\delta=0$ and $B=1-r \cos\theta$ where $r=\sqrt{2\psi}$ and the magnetic field is normalized to its on-axis value \citep{White_84}. In this case, we have $\partial \psi / \partial P_\zeta = 0$ and the original canonical momenta, as defined by Eq. (\ref{canonical}), become $P_\theta=\psi$ and $P_\zeta=\rho_\parallel-\psi_p(P_\theta)$, so that $\bar{P_\zeta}=\rho_\parallel$ and $\bar{\zeta}=\zeta$. The new variables $(\bar{P_\theta},\bar{\theta})$ provide the deviation of the guiding center orbit from a magnetic field line of reference, for which $\theta=\zeta / q(\psi)$, intersecting the poloidal plane $\zeta=0$ at  $(\psi_0,\theta_0)=(P_{\theta 0}, \bar{\theta})$. The new canonical angle $\bar{\theta}$ is directly related to the toroidal precession, defined as $\dot{\zeta}-q\dot{\theta}$. In contrast to previous approaches \citep{Brizard_14a} where an a-posteriori construction of a canonically conjugate pair of variables $(\rho_\parallel, s)$ for the LAR equilibrium case is necessary, within the context of our formulation, the canonical variables are a priori defined. Moreover, this also holds, not only for the LAR case but also, for an arbitrary equilibrium magnetic field.

For a magnetic field configuration that can be considered as a higher order perturbation of the LAR equilibrium, according to the standard tokamak ordering, we can write $\psi=P_\theta+\epsilon^2\hat{\psi}(P_\theta,P_\zeta)$, with $\epsilon$ being the inverse aspect ratio and $\hat{\psi}(P_\theta, P_\zeta)$ corresponding to corrections due to ellipticity and triangularity \citep{White_book}. Therefore, for the new variable $\bar{\zeta}$ we have 
\begin{equation}
\zeta =\bar{\zeta} \left( 1+\epsilon^2\int^{P_\theta}\frac{\partial q^{-1}}{\partial \hat{\psi}}\frac{\partial \hat{\psi}(P'_\theta, P_\zeta)}{\partial P_\zeta} dP'_\theta \right)
\end{equation}
and to lowest order $\bar{\zeta}=\zeta$, whereas the constant of the motion can be written as
\begin{equation}
 P_\zeta(\bar{P_\theta},\bar{P_\zeta})=\bar{P_\zeta}-\psi_p(P_\theta)-\epsilon^2\int^{P_\theta}\frac{\partial q^{-1}(P'_\theta)}{\partial \psi} \psi (P'_\theta,P_\zeta)dP'_\theta \label{nonLAR}
\end{equation}
and to lowest order
\begin{equation}
 \bar{P_\zeta}=P_\zeta+\psi_p(P_{\theta 0}+\bar{P_\theta}).
\end{equation}
For the case of a smooth $q$ profile we can neglect (locally) higher order derivatives of $\psi_p$ with respect to $\psi=P_\theta$, which is equivalent to neglecting the radial variation of the safety factor $q$ (magnetic shear) within a guiding center drift orbit, and applying a Taylor expansion in $\bar{P_\theta}/P_{\theta 0}$ results in
\begin{equation}
 \bar{P_\zeta}=P_\zeta+\psi_p(P_{\theta 0})+\psi'_p(P_{\theta 0})\bar{P_\theta} \label{Pz_Pzbar}
\end{equation}
with prime denoting differentiation of $\psi_p$ with respect to its argument. Since $P_\zeta$ is constant, we can write
\begin{equation}
 \bar{P_\zeta}=\rho_{\parallel 0}+q^{-1}(P_{\theta 0})\bar{P_\theta}. \label{P-P}
\end{equation}
with $\rho_{\parallel 0}=P_{\zeta}+\psi_p(P_{\theta 0})$ depending on both the invariant $P_{\zeta}$ and the flux surface of reference related to $P_{\theta 0}$. This relation imposes a constraint on the two canonical momenta $\bar{P_\zeta}$ and $\bar{P_\theta}$ due to the axisymmetry of the magnetic field and the corresponding invariance of $P_\zeta$. It is worth mentioning that for the case of drift orbits spanning a larger radial distance, we can also include the second order derivative $\psi_p''(P_{\theta 0})$, related to the magnetic shear $(q')$, that would result in a quadratic term with respect to $\bar{P}_\theta$ in (\ref{Pz_Pzbar}) and (\ref{P-P}). As a result the magnetic shear can modify the guiding center dynamics and orbital spectrum \cite{Shaing_15, Albert_16}.

In the new canonical variables the Hamiltonian is given as
\begin{eqnarray}
 H(\bar{P_\zeta},\bar{\zeta}, \bar{P_\theta},\bar{\theta} ; \bar{\mu})&=&\frac{\bar{P_\zeta}^2}{2g^2(\psi(P_{\theta 0}+\bar{P_\theta}))}B^2(\psi(P_{\theta 0}+\bar{P_\theta}), \frac{\bar{\zeta}}{q(\psi(P_{\theta 0}+\bar{P_\theta}))}+\bar{\theta}) \nonumber \\
 & &+\bar{\mu} B(\psi(P_{\theta 0}+\bar{P_\theta}),\frac{\bar{\zeta}}{q(\psi(P_{\theta 0}+\bar{P_\theta}))}+\bar{\theta}) + O(\epsilon^2). \label{H3}
\end{eqnarray}
It is clear that in the new variables there is no cyclic angle, so that neither $\bar{P_\theta}$ nor $\bar{P_\zeta}$ is a constant of the motion. However, the quantity $P_\zeta$ is still a constant of the motion, due to axisymmetry, restricting $\bar{P_\theta}$ and $\bar{P_\zeta}$ according to Eq. (\ref{P-P}), so that integrability is preserved. 


By substituting either of the canonical momenta as a function of the other from (\ref{P-P}), the Hamiltonian system can be readily described in one degree of freedom either as $H(\bar{P_\zeta},\bar{\zeta};\rho_{\parallel 0},P_{\theta 0};\bar{\mu})$ or as $H(\bar{P_\theta},\bar{\theta};\rho_{\parallel 0}, P_{\theta 0}; \bar{\mu})$. In the former case, since the Hamiltonian does not depend on $\bar{P_\theta}$, its canonically conjugate angle $\bar{\theta}$ appears as an additive phase constant that can be omitted, whereas the same holds for $\bar{\zeta}$ in the latter case. This duality directly relates the radial and poloidal drift with the toroidal momentum and precession.

The Hamiltonian (\ref{H3}) provides a canonical description of the guiding center motion for a generic axisymmetric magnetic field equilibrium to the lowest order with respect to the standard tokamak ordering and drift orbit deviation from a magnetic field line. Higher order magnetic field equilibria, with respect to the inverse aspect ratio $\epsilon$,  can be considered by keeping higher order terms in (\ref{nonLAR}). The one degree of freedom Hamiltonian $H(\bar{P_\zeta},\bar{\zeta};\rho_{\parallel 0},P_{\theta 0};\mu)$ can be readily used to calculate the Action 
\begin{equation}
J^{(b,t)}_\zeta=\frac{1}{2 \pi} \oint \bar{P_\zeta}(\bar{\zeta};E,\bar{\mu},\rho_{\parallel 0},P_{\theta 0})d\bar{\zeta}.
\end{equation}
as well as the respective frequencies 
\begin{equation}
\hat{\omega}^{(b,t)}_\zeta=\frac{\partial H}{\partial J^{(b,t)}_\zeta}. 
\end{equation}
with $E=H$ being the constant energy value of each orbit and the index $(b,t)$ corresponding to the cases of trapped (bounce) or passing (transit) orbits.
The mixed-variable generating function \citep{Goldstein} 
\begin{equation}
 \hat{F}(\bar{\zeta},\bar{\theta},\bar{\xi};J_\zeta,J_\theta,J_\xi)=\bar{\xi} J_\xi+\bar{\theta}J_\theta+\int^{\bar{\zeta}} \bar{P}_\zeta(\bar{\zeta}';J_\zeta,J_\theta,J_\xi)d\bar{\zeta}' \label{F}
\end{equation}
provides the canonical transformation to Action-Angle variables also for the remaining canonical variable pairs 
\begin{eqnarray}
 (\bar{P}_\zeta,\bar{\zeta})&\rightarrow&(J_\zeta,\hat{\zeta}) \nonumber \\
 (\bar{P}_\theta,\bar{\theta})&\rightarrow&(J_\theta,\hat{\theta})\\
 (\bar{\mu},\bar{\xi})&\rightarrow&(J_\xi,\hat{\xi}) \nonumber
\end{eqnarray}
It is worth emphasizing that in the other two pairs of canonical variables, the new momenta (Actions) are identical to the old momenta, i.e. $J_\theta=\bar{P}_\theta$ and $J_\xi=\bar{\mu}=\mu$ but the new positions (angles) differ from the old ones, due to the last term of the generating function (\ref{F}) \citep{Kaufman_72}.    
The transformation to Action-Angle variables allows for the calculation of the orbital frequencies in the remaining two degrees of freedom. Therefore, 
\begin{equation}
 \hat{\omega}_\theta=\frac{\partial H}{\partial J_\theta}=-\frac{\partial H}{\partial J_\zeta}\frac{\partial J_\zeta}{\partial J_\theta}
\end{equation}
and 
\begin{equation}
 \hat{\omega}_\xi=\frac{\partial H}{\partial J_\xi}=-\frac{\partial H}{\partial J_\zeta}\frac{\partial J_\zeta}{\partial J_\xi}
\end{equation}
or
\begin{equation}
 \frac{\hat{\omega}_\theta}{\hat{\omega}_\zeta^{(b,t)}}=-\frac{\partial J_\zeta^{(b,t)}}{\partial J_\theta} \label{omega_theta}
\end{equation}
and 
\begin{equation}
  \frac{\hat{\omega}_\xi}{\hat{\omega}_\zeta^{(b,t)}}=-\frac{\partial J_\zeta^{(b,t)}}{\partial J_\xi}. \label{omega_xi}
\end{equation}
The above equations show that the bounce/transit Actions $J_\zeta^{(b,t)}$ (with a sign change) can be considered as the Hamiltonian functions providing the canonical equations for the other degrees of freedom when time is normalized with respect to the inverse bounce/transit frequency $\hat{\omega}_\zeta^{(b,t)}$ \citep{White_book}. 
Along with the analytical expressions obtained in the next Section, these equations provide compact formulas for bounce/transit-averaged dynamics under low-frequency electromagnetic fluctuations in the context of a bounce-gyrokinetic theory \citep{Brizard_00, Brizard_14b}. 

The frequency $\hat{\omega}_\theta$ is directly related to the bounce/transit averaged toroidal drift as shown in the following expression
\begin{eqnarray}
\frac{\hat{\omega}_\theta}{\hat{\omega}_\zeta}&=&-\frac{1}{2 \pi}\oint \frac{\partial}{\partial J_\theta} \bar{P_\zeta}(\bar{\zeta};E,\bar{\mu},\rho_{\parallel 0},P_{\theta 0})d\bar{\zeta}=\frac{1}{2 \pi}\oint \frac{\partial \bar{P_\zeta}}{\partial E} \frac{\partial E}{\partial J_\theta} d\bar{\zeta} \nonumber \\
&=&\frac{1}{2 \pi}\oint \left(\frac{d\bar{\zeta}}{dt}\right)^{-1} \frac{\partial E}{\partial \bar{P}_\theta} d\bar{\zeta}=\frac{1}{2 \pi}\oint \left(\frac{d\bar{\theta}}{dt}\right) dt=\frac{1}{2 \pi} \left(\Delta \bar{\theta}\right)_{T_\zeta=2\pi/\hat{\omega}_\zeta} \label{num_precession}
\end{eqnarray}
where $\left(\Delta \bar{\theta} \right)_{T_\zeta=2 \pi/ \hat{\omega}_\zeta}$ is the variation of $\bar{\theta}=\theta-\zeta/q$ in the time interval of a period $T_\zeta=2\pi/\hat{\omega}_\zeta$.
Similarly, the frequency $\hat{\omega}_\xi$ corresponds to the bounce/transit averaged gyrofrequency \citep{Kaufman_72}
\begin{equation}
\frac{\hat{\omega}_\xi}{\hat{\omega}_\zeta}= \frac{1}{2 \pi} \left(\Delta \bar{\xi} \right)_{T_\zeta=2\pi/\hat{\omega}_\zeta}.
\end{equation}


The Zero Drift Width (ZDW) approximation, under which the guiding center is considered as being fixed on a given flux surface $P_{\theta 0}=\psi_0$, can be readily obtained by setting $\bar{P_\theta}=0$. Moreover, for a relatively small drift width $\bar{P_\theta}<<P_{\theta 0}$, $\psi$ can be Taylor-expanded around $P_{\theta 0}$ and with $\bar{P_\theta}$ substituted from (\ref{P-P}) we have a single degree of freedom Hamiltonian for the canonical variables $(\bar{P_\zeta},\bar{\zeta})$.  

For a LAR equilibrium the Full Drift Width (FDW) Hamiltonian is written as 
\begin{equation}
 H_{FDW}(\rho_\parallel,\zeta; \rho_{\parallel 0}, \mu; \psi_0)=\frac{\rho_\parallel^2}{2}\left[1-\sqrt{2\psi} \cos\left(\frac{\zeta}{q\left( \psi \right)} \right) \right]^2+\mu  \left[1-\sqrt{2\psi} \cos\left(\frac{\zeta}{q\left( \psi \right)} \right) \right] \label{H4}
\end{equation}
where we have substituted $\bar{P_\zeta}=\rho_\parallel$ and dropped bar in $\zeta$ for simplicity. It is worth emphasizing that obtaining the Hamiltonian (\ref{H4}) is enabled by the utilization of the normalized parallel guiding-center velocity $\rho_\parallel$ (normalized to the magnetic field), instead of the regular parallel guiding-center velocity used in previous works \cite{Brizard_11}. Moreover, according to Eq. (\ref{P-P}), $\psi$ is taken as a function of $\rho_\parallel$ from 
\begin{equation}
\psi= \psi_0 +q(\psi_0)(\rho_\parallel-\rho_{\parallel 0}) 
\end{equation}
with $\psi_0=P_{\theta 0}$ and $\rho_{\parallel 0}=P_{\zeta 0}+\psi_p(\psi_0)$. 

\section{Zero Drift Width Approximation}
We consider the Zero Drift Width approximation under which the guiding center motion is considered fixed on a given flux surface $\psi=\psi_0$ and the frequencies of $\theta$ and $\zeta$ are simply related as $\omega_\zeta=q(\psi_0)\omega_\theta$. The Hamiltonian is written as
\begin{equation}
 H_{ZDW}(\rho_\parallel,\zeta; \psi_0, \mu)=\frac{\rho_\parallel^2}{2}\left[1-\sqrt{2\psi_0} \cos\left(\frac{\zeta}{q\left( \psi_0 \right)} \right) \right]^2+\mu  \left[1-\sqrt{2\psi_0} \cos\left(\frac{\zeta}{q\left( \psi_0 \right)} \right) \right] \label{H_ZDW}
\end{equation}
or equivalently
\begin{equation}
 H_{ZDW}(\rho_\parallel,\zeta; \psi_0, \mu)=\frac{\rho_\parallel^2}{2}+\mu\left\{1-\left[ 1+ \frac{\rho_\parallel^2}{\mu} -\frac{\rho_\parallel^2}{2\mu} \sqrt{2\psi_0}  \cos\left(\frac{\zeta}{q\left( \psi_0 \right)} \right) \right] \sqrt{2\psi_0}  \cos\left(\frac{\zeta}{q\left( \psi_0 \right)} \right) \right\}
 \label{H_ZDW_order}
\end{equation}
Further approximations can be based on the relative magnitude of the three terms in the square brackets of the above equation. For $\rho_\parallel^2 /\mu << 1$, corresponding to particles with large pitch angles $\alpha=\tan^{-1}(v_\perp/|v_\parallel|)$, the Hamiltonian is reduced to
\begin{equation}
 H'_{ZDW}(\rho_\parallel,\zeta; \psi_0, \mu)=\frac{\rho_\parallel^2}{2}+\mu \left[1-\sqrt{2\psi_0}  \cos\left(\frac{\zeta}{q\left( \psi_0 \right)} \right)\right]  \label{H_ZDW_prime}
\end{equation}
which resembles the Hamiltonian of a pendulum, with the trapped and passing orbits corresponding to a libration and rotation type of motion \citep{Shaing_09, Brizard_11, Brizard_14a}. It is worth mentioning that, under the ZDW approximation, both $H_{ZDW}$ and $H'_{ZDW}$ scale with $\mu$ when $\rho_\parallel^2$ is also divided by $\mu$, which clearly is not the case for the FDW Hamiltonian [Eq.(\ref{H4})].  

 The above approximations have significant quantitative and qualitative differences: They provide different values for the total energy of a specific orbit and also result in different separatrices betweeen trapped and passing orbits in the phase space, as shown in Fig. 1. Therefore, a particle that is described as being trapped according to $H'_{ZDW}$ can be actually passing according to $H_{ZDW}$ and vice versa. Both ZDW Hamiltonians describe guiding center orbits that are symmetric with respect to $\rho_\parallel=0$ whereas this is not the case with the FDW Hamiltonian, according to which positive (co-passing) and negative (counter-passing) orbits are not symmetric. These differences are not uniform across the phase space of the system and depend strongly on the pitch angle and the flux surface of reference $\psi_0$. 

The ZDW Hamiltonians are very useful for obtaining analytical forms for the frequencies of the guiding center motion and their dependence on the constants of the motion, parametrizing each orbit. These frequencies determine the Orbital Spectrum (OS) of different particle species, including bulk and energetic particles, as well as their resonance conditions with any type of non-axisymmetric perturbations. The latter are crucial for the energy, momentum and particle transport in a toroidal magnetic field configuration. 

For both $H_{ZDW}$ and $H'_{ZDW}$ bounce and transit motion is characterized by $0 \leq k < 1$ and $1 < k$, respectively, with $k$ is the trapping parameter, defined as
\begin{equation}
 k=\frac{E-\mu(1-r)}{2\mu r},
\end{equation}
and 
\begin{equation}
 r=\sqrt{2\psi_0}.
\end{equation}

The bounce/transit frequencies and Actions corresponding to $H'_{ZDW}$ are given as follows (Appendix).
\begin{eqnarray}
\omega'_b&=&\frac{\pi \sqrt{\mu r}}{2 q(\psi_0) K(k)}=\frac{\pi}{2K(k)}\omega'_{b0}, \label{omega_b_prime}\\ 
\omega'_t&=&\frac{\pi \sqrt{\mu r}\sqrt{k}}{q(\psi_0)K(k^{-1})}=\frac{\pi\sqrt{k}}{K(k^{-1})}\omega'_{b0},\label{omega_t_prime} 
\end{eqnarray}
and
\begin{eqnarray}
J'_b&=&\frac{8q(\psi_0)\sqrt{\mu r}}{\pi}[E(k)+(k-1)K(k)] , \label{J_b_prime}\\
J'_t&=&\frac{4q(\psi_0)\sqrt{\mu r}}{\pi}\sqrt{k}E(k^{-1}) \label{J_t_prime}
\end{eqnarray}
where $K$ and $E$ are the complete elliptic integrals of the first and second kind \citep{EllipticInt}, and 
\begin{equation}
 \omega'_{b 0}=\frac{\pi \sqrt{\mu r}}{2 q(\psi_0) K(0)}=\frac{ \sqrt{\mu r}}{q(\psi_0)}
\end{equation}
 is the frequency of the deeply trapped bounce motion, corresponding to $k=0$. These are the standard analytical expressions for the frequencies and Actions obtained from the pendulum-like Hamiltonian $H'_{ZDW}$ \citep{White_book,  Shaing_09, Brizard_11}.

The respective frequencies and Actions for $H_{ZDW}$ are (Appendix)
\begin{eqnarray}
\omega_b&=&\frac{\pi (1-r)\sqrt{\mu r}}{2 q(\psi_0)\Pi\left(\eta k, k\right)}=\frac{\pi (1-r)}{2 \Pi\left(\eta k, k\right)}\omega'_{b 0}, \label{omega_b}\\
\omega_t&=&\frac{\pi \sqrt{k} (1-r)\sqrt{\mu r}}{q(\psi_0)\Pi\left(\eta, k^{-1}\right)}=\frac{\pi \sqrt{k} (1-r)}{\Pi\left(\eta, k^{-1}\right)}\omega'_{b 0}, \label{omega_t}
\end{eqnarray}
and 
\begin{eqnarray}
 J_b&=&\frac{8 q(\psi_0)\sqrt{\mu r}}{\pi \eta (1-r)} \left[ (\eta k-1)\Pi\left(\eta k,k\right)+ K(k) \right], \label{J_b}\\
J_t&=&\frac{4 q(\psi_0)\sqrt{\mu r}}{\pi \eta (1-r)}\left[ \frac{\eta k-1}{\sqrt{k}}\Pi\left(\eta,k^{-1}\right)+ \frac{K(k^{-1})}{\sqrt{k}} \right], \label{J_t}
 \end{eqnarray}
with $\Pi$ being the complete elliptic integral of the third kind \citep{EllipticInt},
and
\begin{equation}
 \eta=-\frac{2r}{1-r}.
\end{equation}


For a given set of values of the magnetic moment $(\mu)$ and the flux surface $(\psi_0)$, the value of the trapping parameter $(k)$ is determined by the total energy $(E)$ of the particle. The dependence of the bounce and transit frequencies on the energy $E(k)$ according to $H'_{ZDW}$ and $H_{ZDW}$, as given by Eqs. (\ref{omega_b_prime})-(\ref{omega_t_prime}) and Eqs. (\ref{omega_b})-(\ref{omega_t}) respectively, is depicted in Figs. 2 and 3. The bounce and transit frequencies $\dot{\theta}$ according to the Full Drift Width Hamiltonian in the original canonical variable set, given in Eq. (\ref{H2}), are also shown. Note that under the ZDW approximation $\dot{\theta}=\dot{\zeta}/q=\dot{\bar{\zeta}}/q=\omega_{b,t}/q=\hat{\omega}_\zeta/q$, and we have taken $q=1$ for simplicity. The analytical formula for $\omega_b$ (\ref{omega_b}) shows a remarkable agreement with the numerically calculated frequencies based on the FDW Hamiltonian and significantly deviate from the analytical formula for $\omega'_b$ (\ref{omega_b_prime}) with the deviation being more pronounced for larger values of $\psi_0$, as shown in Fig. 2. For the transit frequencies, shown in Fig. 3, the FDW Hamiltonian describes asymmetric orbits and consequently different frequencies for co-passing $(\rho_\parallel>0)$ and counter-passing $(\rho_\parallel<0)$ orbits. The analytical formula for $\omega_t$ (\ref{omega_t}) corresponds to an intermediate frequency with respect to the two branches of the numerically calculated FDW frequencies, whereas the analytical formula for  $\omega'_t$ (\ref{omega_t_prime}) tends to follow one of the two branches.

The two analytical formulas for the bounce freqencies have also a significant qualitative difference. In fact, in contrast to $\omega'_b$, $\omega_b$ is non-monotonic with respect to $k$ (energy) and, in accordance to FDW numerical calculations, predicts a local maximum of the frequency of the trapped orbits with respect to the energy, indicating the existence of two trapped orbits with different energy but equal frequency, as shown in Fig. 2(b). The location of this local maximum in the space $(E(k), \psi_0)$ is depicted in Fig. 3. 
The nonmonotonic dependence of the frequency on the particle energy has important implications for particle and momentum transport under the presence of non-axisymmetric perturbations that break the integrability of the system. In such cases, the locations of the phase space regions where orbits are significantly modified due to perturbations are determined by resonance conditions. The nonmonotonicity suggests that the same resonance can modify two regions of the phase space leading to extended transport under resonance overlap conditions. Moreover, the existence of energy values where the derivative of the frequency with respect to energy is zero can be related to an intrinsic degeneracy condition which is crucial for the effect of perturbations \citep{Li-Li}.   

The bounce-averaged toroidal precession and gyration frequencies can be  analytically caclulated according to Eqs. (\ref{omega_theta}) and (\ref{omega_xi}) with the utilization of either $J_b$ (\ref{J_b}), $\omega_b$ (\ref{omega_b}) or $J'_b$ (\ref{J_b_prime}), $\omega'_b$ (\ref{omega_b_prime}). The resulting expressions are too lengthy to be given here; their dependence on energy $(k)$ is depicted in Figs. 5 and 6. A sign reversal in the bounce-averaged toroidal precession frequency is shown in Fig. 5. The analytical results based on $H_{ZDW}$ show a remarkable agreement with the numerical results based on the FDW Hamiltonian, where the expression (\ref{num_precession}) has been utilized. The differences between analytical results based on $H_{ZDW}$ and $H'_{ZDW}$ significantly differ for larger values of $\psi_0$. These differences imply different phase space regions where resonant interactions with low-frequency electromagnetic perturbations take place \citep{Brizard_14b}. Similar differences are also show for the bounce-averaged gyration frequencies, in Fig. 6, determining the conditions for resonant interactions with high-frequency waves.

\section{Summary and Conclusions}
The guiding center motion in an axisymmetric magnetic field is analyzed under a Hamiltonian formulation in canonical variables. The Zero Drift Width approximation has been described in the context of the canonical formulation through a canonical transformation to variables measuring the deviation of the guiding center from a magnetic field line of reference. The latter allows for the expression of the system in Action-Angle variables and the systematic dynamical reduction to lower dimensional phase spaces, depending on the specific physical problem under consideration, such as low or high frequency perturbations.

A novel Zero Drift Width Hamiltonian has been obtained along with compact analytical formulas for all the guiding center Orbital Frequencies, namely bounce/transit frequency and bounce-averaged toroidal precession and gyration frequency. The analytical results significantly differ from those corresponding to the widely used pendulum-like Hamiltonian and show a remarkable agreement with numerically calculated frequencies for the Full Drift Width Hamiltonian. 

The knowledge of the Orbital Frequencies is crucial for determining the resonance conditions under particle interaction with non-axisymmetric perturbations that affect energy, momentum and particle transport in toroidal plasma configurations. Moreover, the transformation to Action-Angle variables allows for the application of standard canonical perturbation methods as well as the systematic dynamical reduction and the formulation of a bounce gyrkinetic description.

\section*{Acknowledgements}
GA and YK aknowledge useful discussions with Panagiotis A. Zestanakis.  This work has been carried out within the framework of the EUROfusion Consortium and has received funding from the Euratom research and training programme 2014-2018 and 2019-2020 under grant agreement No 633053 as well as from the National Programme for the Controlled Thermonuclear Fusion, Hellenic Republic. The views and opinions expressed herein do not necessarily reflect those of the European Commission.

\section*{Appendix: Action-Angle Variables} 
The Zero Drift Width Hamiltonians $H_{ZDW}$ and $H'_{ZDW}$ given in Eqs. (\ref{H_ZDW}) and (\ref{H_ZDW_prime}), can be written as 
\begin{equation}
 H(\rho_\parallel,\zeta; \psi_0, \mu;\sigma)=\frac{\rho_\parallel^2}{2}\left[1-\sigma r \cos\left(\frac{\zeta}{q\left( \psi_0 \right)} \right) \right]^2+\mu  \left[1-r \cos\left(\frac{\zeta}{q\left( \psi_0 \right)} \right) \right] =E
\end{equation}
with $\sigma=1$ and $\sigma=0$, respectively. The equation $H=E$, where $E$ is the constant energy, can be solved with respect to $\rho_\parallel$ as follows
\begin{equation}
 \rho_\parallel^2=2\frac{(E-\mu)+\mu r \cos\left(\frac{\zeta}{q\left( \psi_0 \right)}\right)}{\left[1-\sigma r \cos\left(\frac{\zeta}{q\left( \psi_0 \right)}\right)\right]^2}
\end{equation}
The corresponding Action variables are defined as
\begin{equation}
 J=\oint\rho_\parallel d\zeta=\oint \frac{\sqrt{2[(E-\mu)+\mu r \cos(\zeta/q)]}}{1-\sigma r \cos(\zeta/q)} d\zeta
\end{equation}
and, by setting $\zeta/2q \equiv\chi$ and using $\cos2\chi=1-2\sin^2\chi$, the Action variables are given as
\begin{equation}
 J=\frac{2^{l+1}q(\psi_0)\sqrt{k}\sqrt{\mu r}}{\pi(1-\sigma r)}\int_{\chi_{min}}^{\chi_{max}}\frac{\sqrt{1-k^{-1}\sin^2\chi}}{1-\eta\sin^2\chi}d\chi
\end{equation}
with $k=[E-\mu(1-r)]/2\mu r$, $\eta=-2\sigma r/(1-\sigma r)$ and $\chi_{min,max}=\sin^{-1}\left(\mp \sqrt{k}\right)$ for trapped orbits (bounce motion, $0\leq k <1$, $l=1$) and $\chi_{min,max}=\mp \pi/2$ for passing orbits (transit motion $k>1$, $l=0$). 

The equation of motion $d\zeta/dt=\partial H / \partial \rho_\parallel$ can be written as
\begin{equation}
 dt=\pm\frac{q(\psi_0)}{(1-\sigma r)\sqrt{k}\sqrt{\mu r}}\frac{d\chi}{\left( 1-\eta\sin^2\chi \right)\sqrt{1-k^{-1}\sin^2\chi}}. \label{dt}
\end{equation}
 The integration of the above relation over a closed orbit provides the period of the bounce and the transit motion as
\begin{equation}
T=\frac{2^lq(\psi_0)}{(1-\sigma r)\sqrt{k}\sqrt{\mu r}} \int_{\chi_{min}}^{\chi_{max}}\frac{d\chi}{\left(1-\eta\sin^2\chi\right)\sqrt{1-k^{-1}\sin^2\chi}}
\end{equation}
with the respective frequency given as $\omega=2\pi/T$. 

The resulting expressions for the bounce and transit frequencies and Actions for the Hamiltonian $H'_{ZDW}$ are given in terms of complete elliptic integrals of the first and second kind \citep{EllipticInt} as in Eqs. (\ref{omega_b_prime}), (\ref{omega_t_prime}) and Eqs. (\ref{J_b_prime}), (\ref{J_t_prime}). For the Hamiltonian $H_{ZDW}$ the corresponding expressions are given in terms of complete elliptic integrals of the third kind \citep{EllipticInt} as in Eqs. (\ref{omega_b}), (\ref{omega_t}) and Eqs. (\ref{J_b}), (\ref{J_t}). 

The equation of motion (\ref{dt}) can also be used for the calculation of the transformation to the Angle variable $\hat{\zeta}$ in terms of incomplete elliptic integrals as follows
\begin{equation}
 \hat{\zeta}\equiv \omega t= \frac{q(\psi_0)\omega(k,\mu,r;\sigma)}{(1-\sigma r)\sqrt{k}\sqrt{\mu r}} \int_{\chi_{min}}^{\chi}\frac{d\chi'}{\left(1-\eta\sin^2\chi'\right)\sqrt{1-k^{-1}\sin^2\chi'}}
\end{equation}
where we have taken $\hat{\zeta}=0$ when $\chi=\chi_{min}$. Note that this relation defines the transformation from $\zeta=2q(\psi_0)\chi$ to $\hat{\zeta}$ in an implicit form. For the case of the Hamiltonian $H'_{ZDW}$ corresponding to $\sigma=0$, the relation can be inverted with the use of Jacobi elliptic functions, whereas for the case of $H_{ZDW}$ corresponding to $\sigma=1$, the inversion of the relation, to the best of our knowledge, cannot be expressed in a convenient form \citep{EllipticInt}.

\clearpage

\clearpage

\begin{figure*}\centering
\begin{center}
     \subfigure[]
       {\includegraphics[width=0.49\columnwidth]{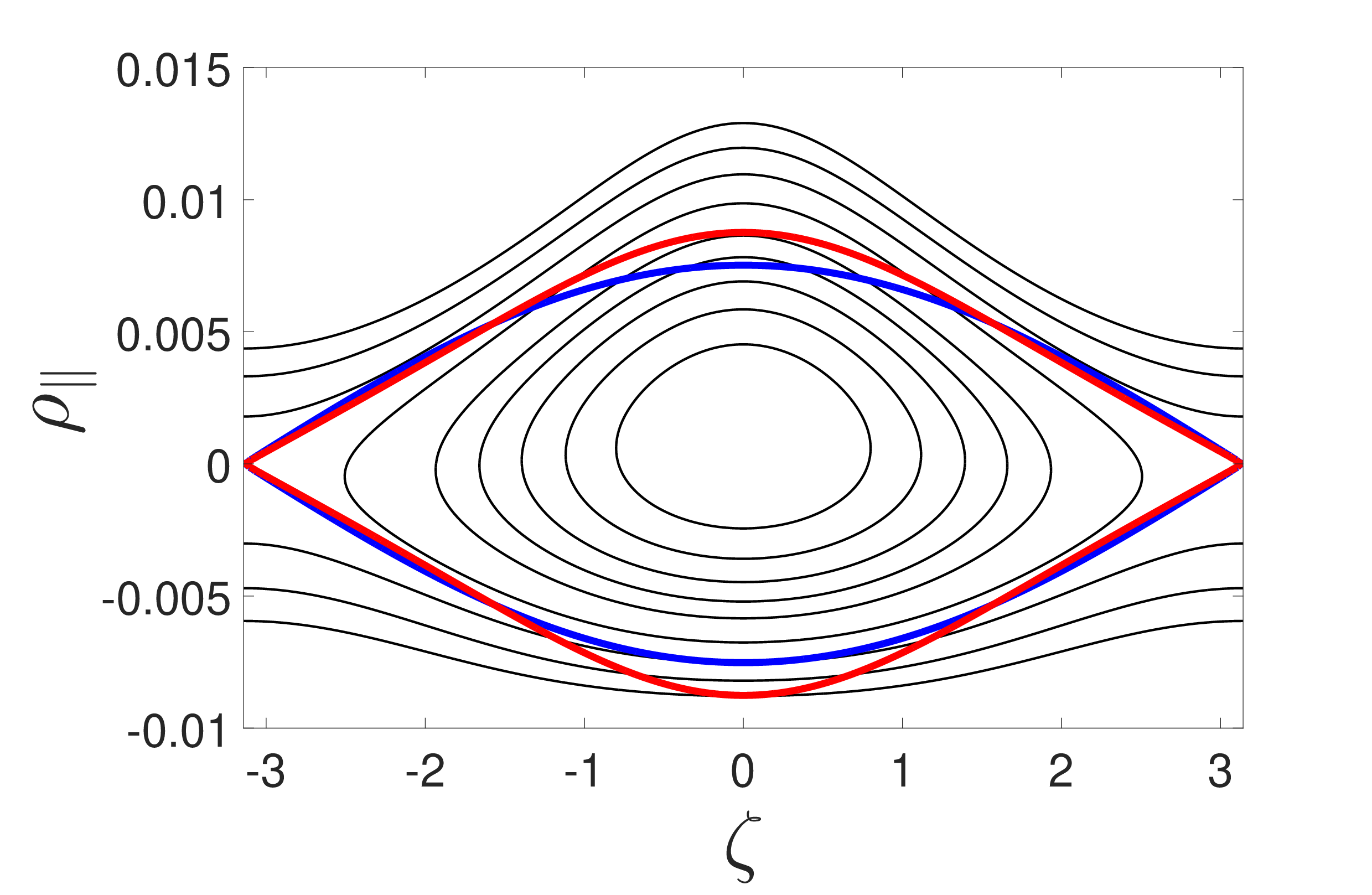}}
     \subfigure[]
          {\includegraphics[width=0.49\columnwidth]{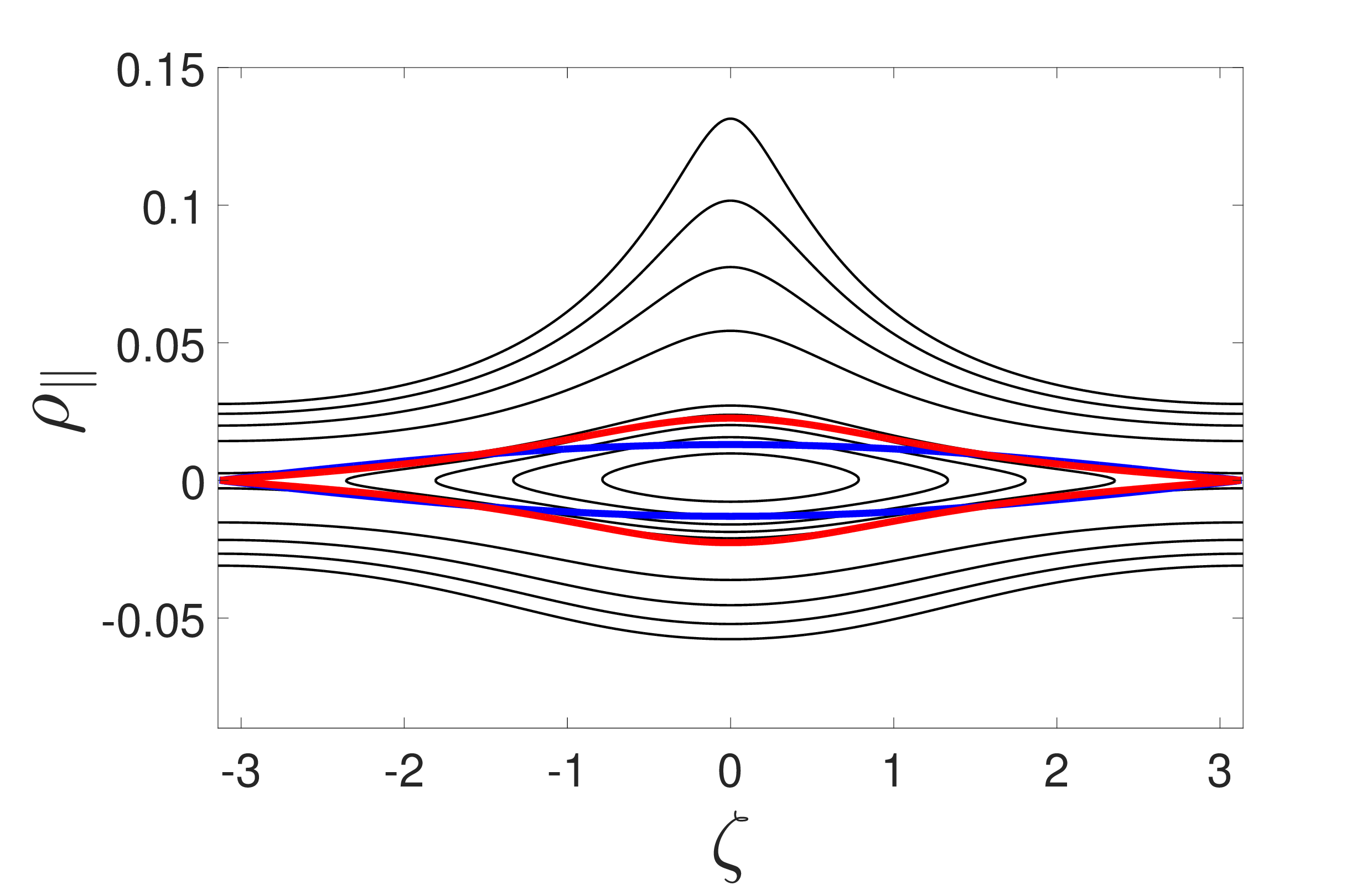}}
     \caption{Phase space $(\rho_\parallel,\zeta)$ of the  Hamiltonian $H_{FDW}$ (\ref{H4}) for $q=1$, $\mu=10^{-4}$, and (a) $\psi_0=0.01$, (b) $\psi_0=0.09$. The separatrices between bounce and transit motion according to $H_{ZDW}$ (\ref{H_ZDW}) and $H'_{ZDW}$ (\ref{H_ZDW_prime}) are depicted by blue and red lines, respectively.}
     \label{fig:ImageSize.testing.quarter}    
\end{center}
\end{figure*}

\begin{figure*}\centering
\begin{center}
     \subfigure[]
       {\includegraphics[width=0.49\columnwidth]{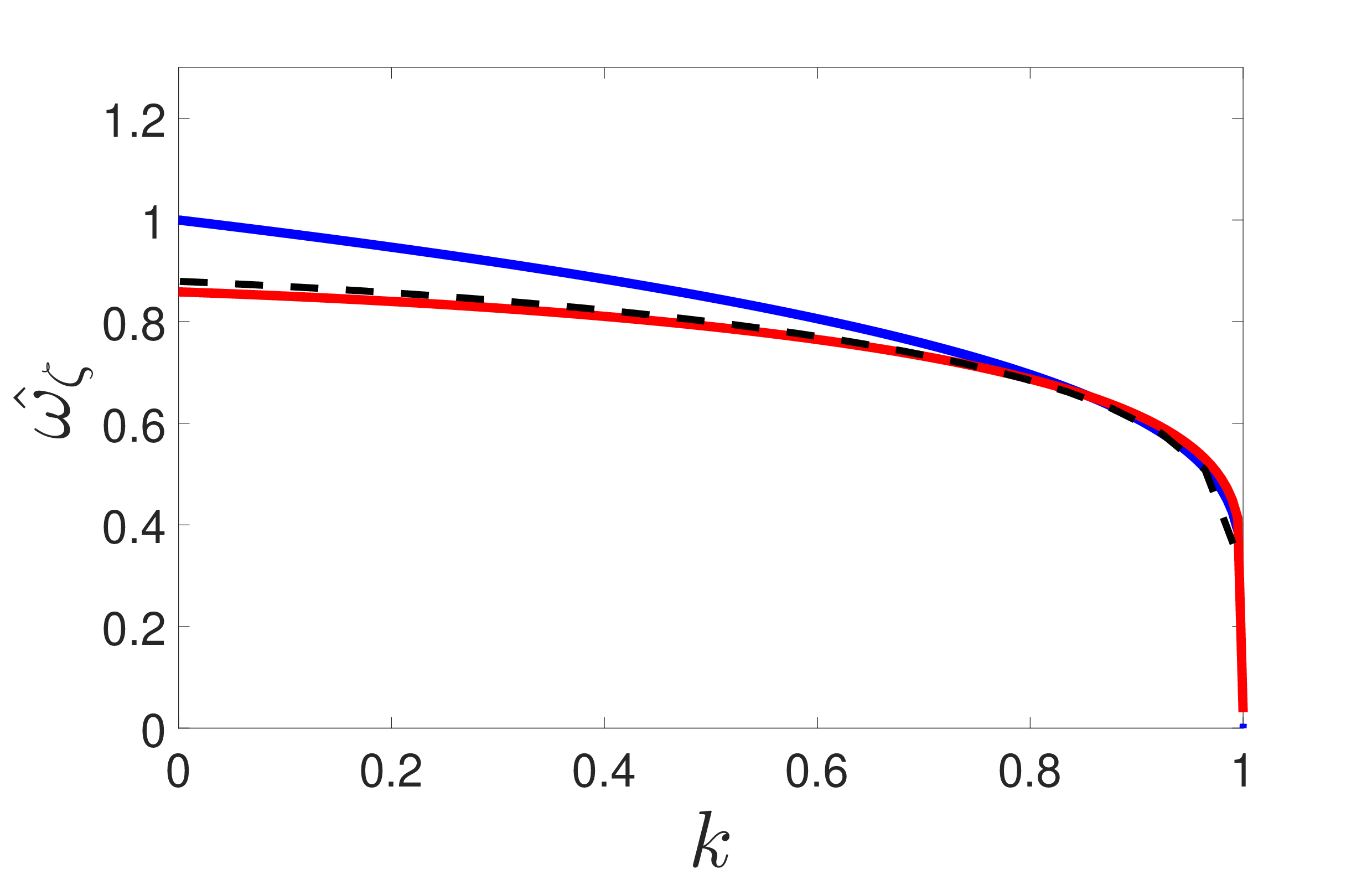}}
     \subfigure[]
          {\includegraphics[width=0.49\columnwidth]{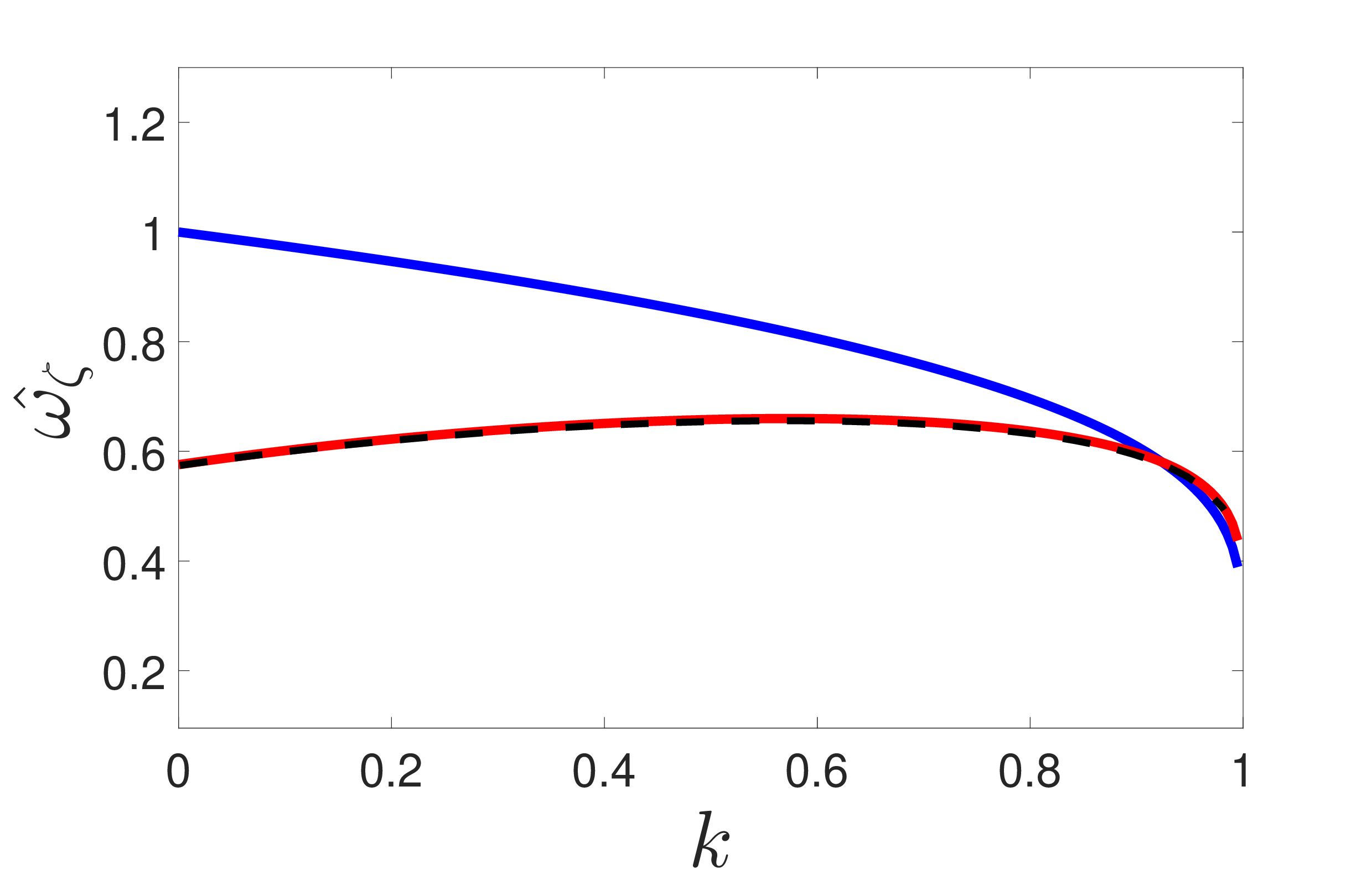}}
     \caption{Analytically (solid lines) and numerically (dashed lines) calculated bounce frequencies according to $H'_{ZDW}$ (blue line), $H_{ZDW}$ (red line) and $H$ [Eq. (\ref{H2})] (dashed line) for $q=1$, $\mu=10^{-4}$, and (a) $\psi_0=0.01$ $(\eta=-0.33)$, (b) $\psi_0=0.09$ $(\eta=-1.47)$.}
     \label{fig:ImageSize.testing.quarter}    
\end{center}
\end{figure*}

\begin{figure*}\centering
\begin{center}
     \subfigure[]
       {\includegraphics[width=0.49\columnwidth]{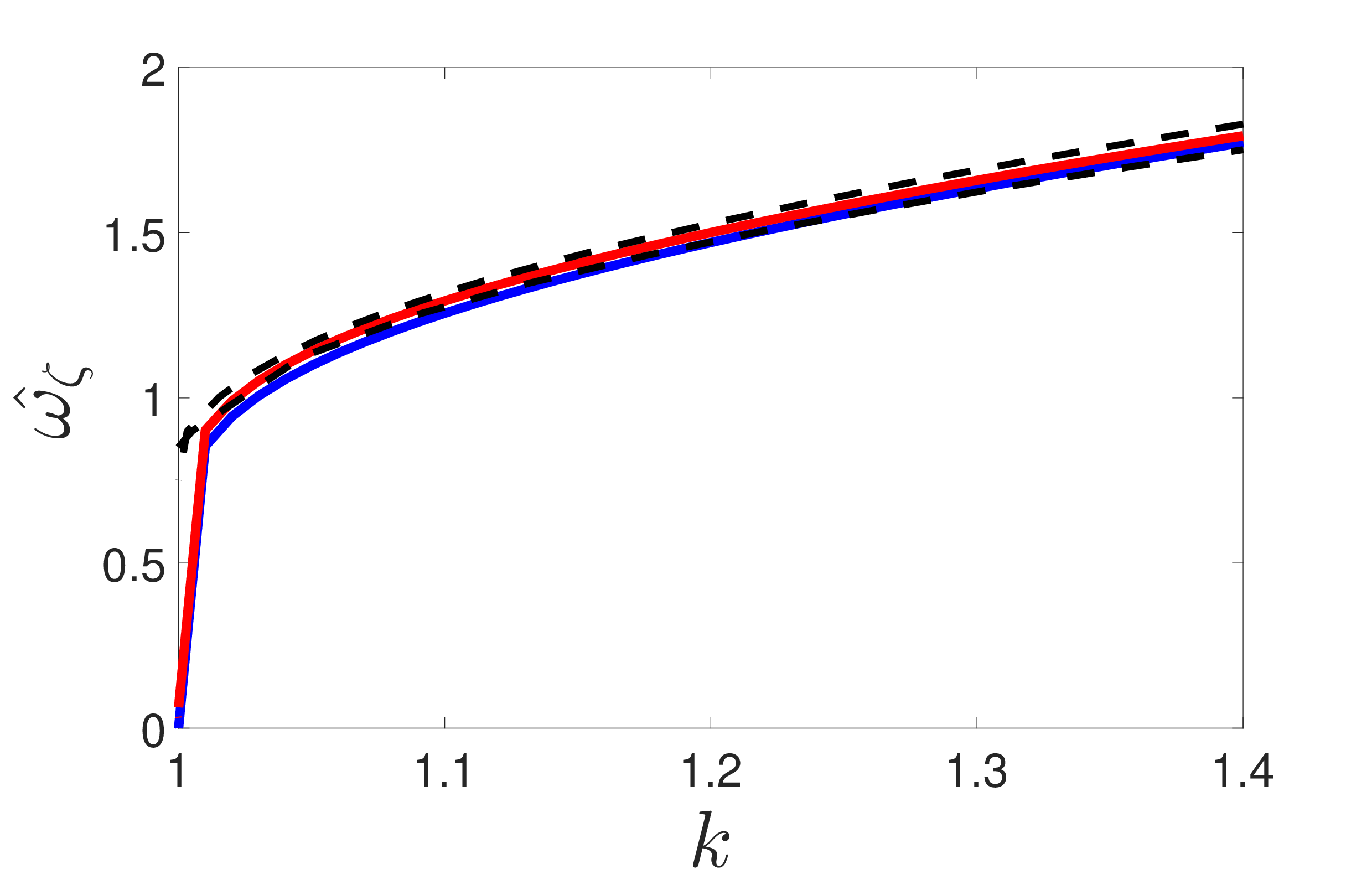}}
     \subfigure[]
          {\includegraphics[width=0.49\columnwidth]{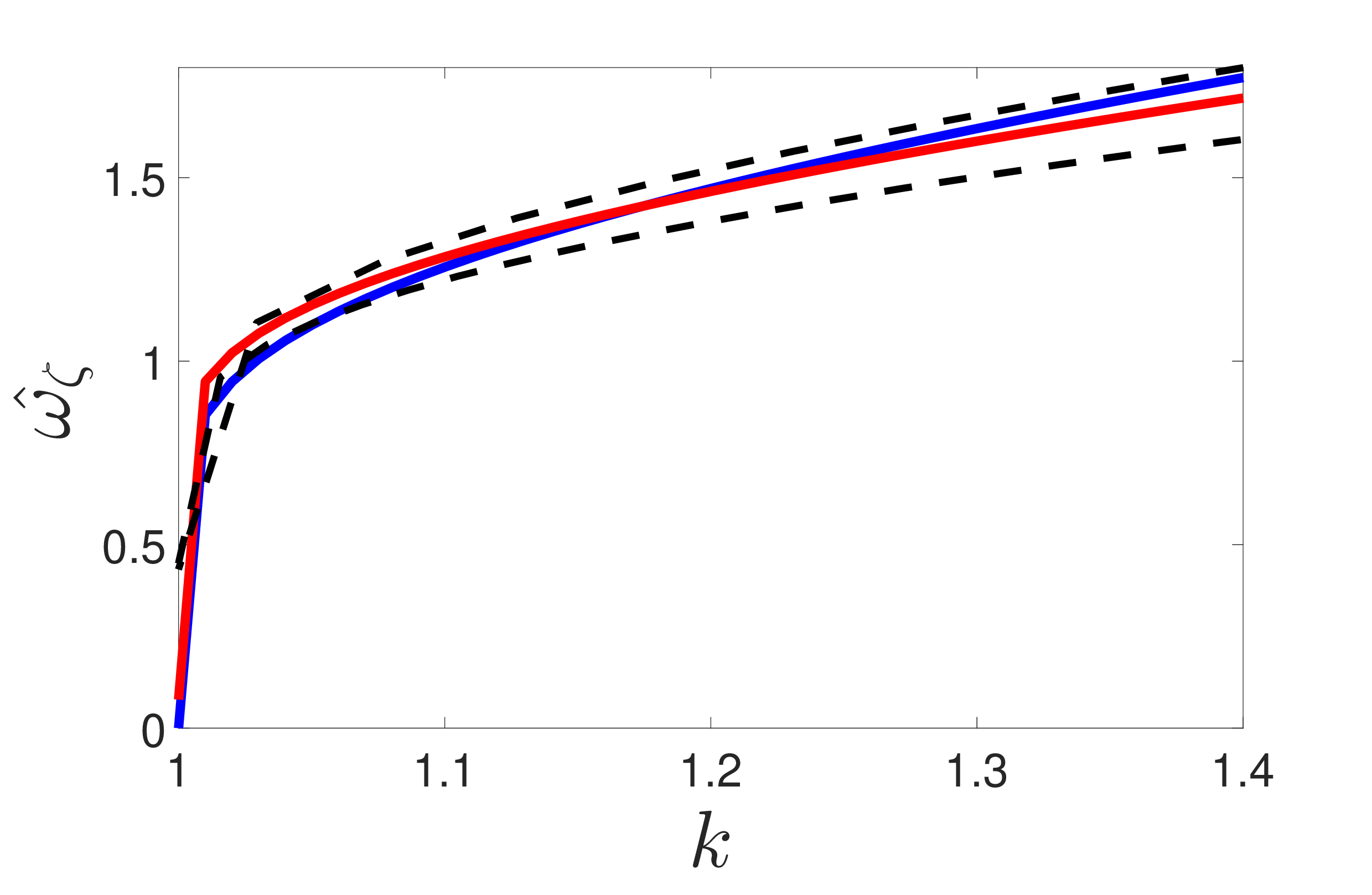}}
     \caption{Analytically (solid lines) and numerically (dashed lines) calculated transit frequencies according to $H'_{ZDW}$ (blue line), $H_{ZDW}$ (red line) and $H$ [Eq. (\ref{H2})] (dashed line) for $q=1$, $\mu=10^{-4}$, and (a) $\psi_0=0.01$ $(\eta=-0.33)$,  (b) $\psi_0=0.09$ $(\eta=-1.47)$.}
     \label{fig:ImageSize.testing.quarter}    
\end{center}
\end{figure*}

\begin{figure*}\centering
\begin{center}
     \subfigure[]
          {\includegraphics[width=0.49\columnwidth]{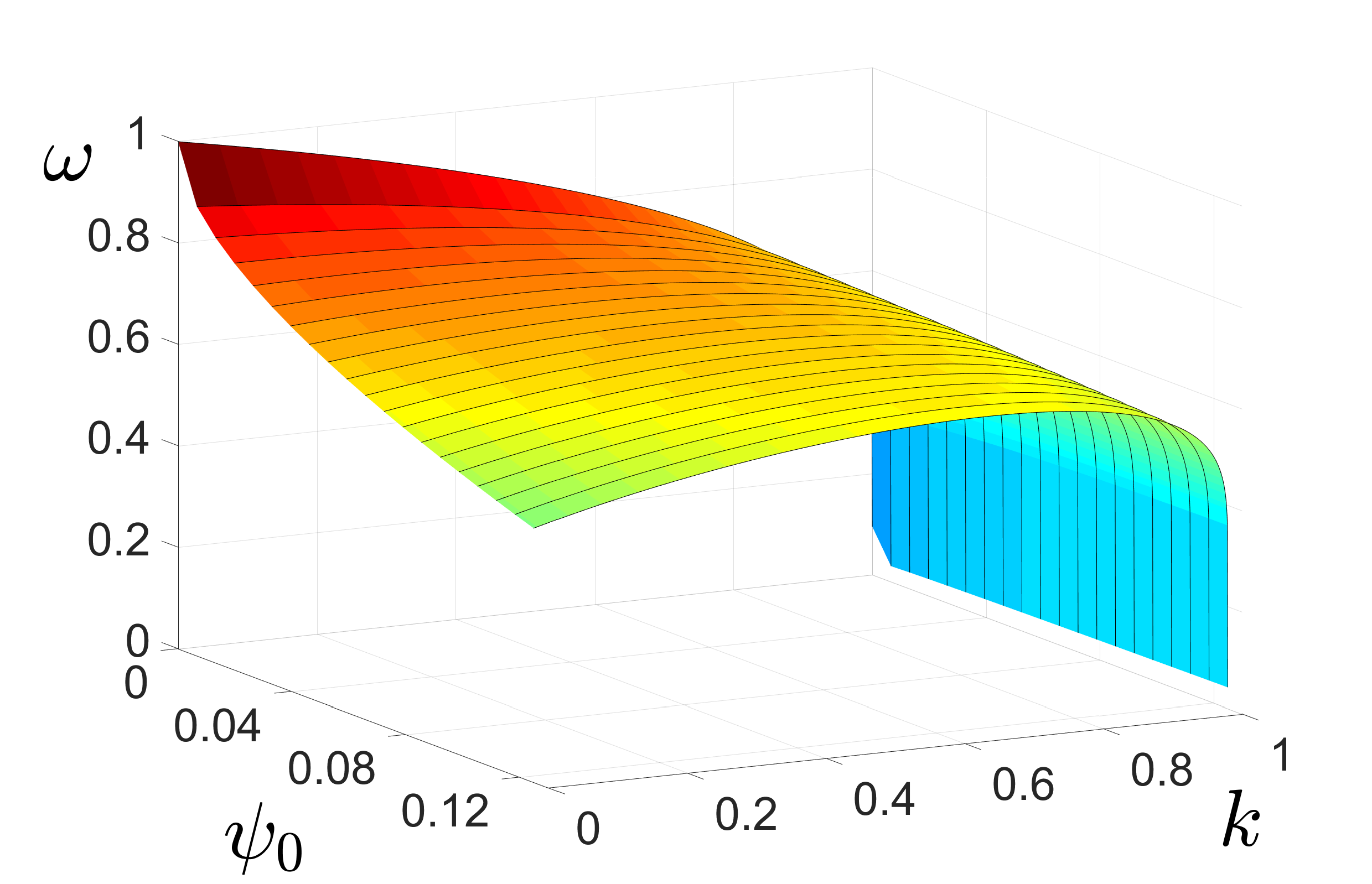}}
     \subfigure[]
          {\includegraphics[width=0.49\columnwidth]{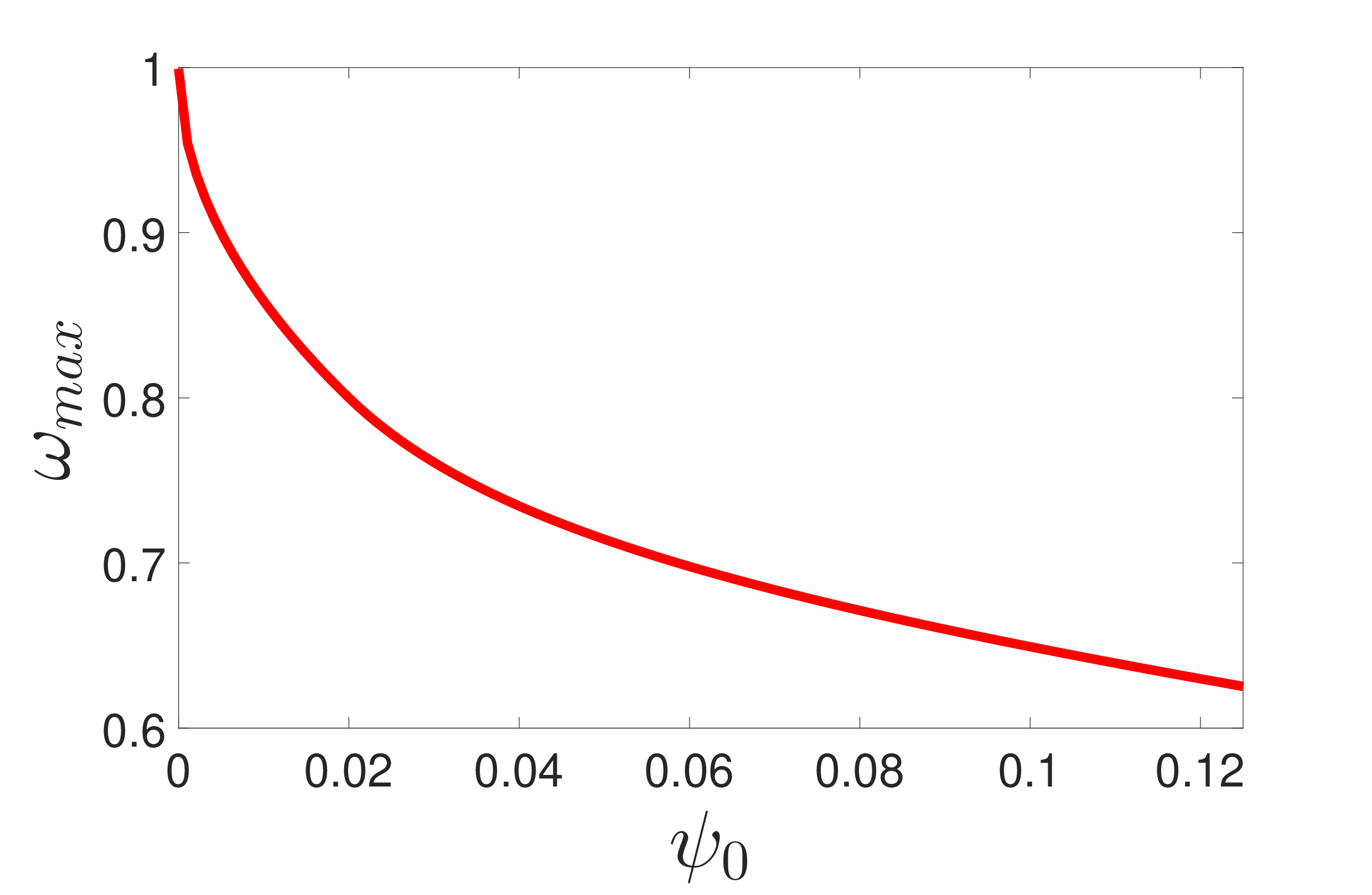}}
     \subfigure[]
          {\includegraphics[width=0.49\columnwidth]{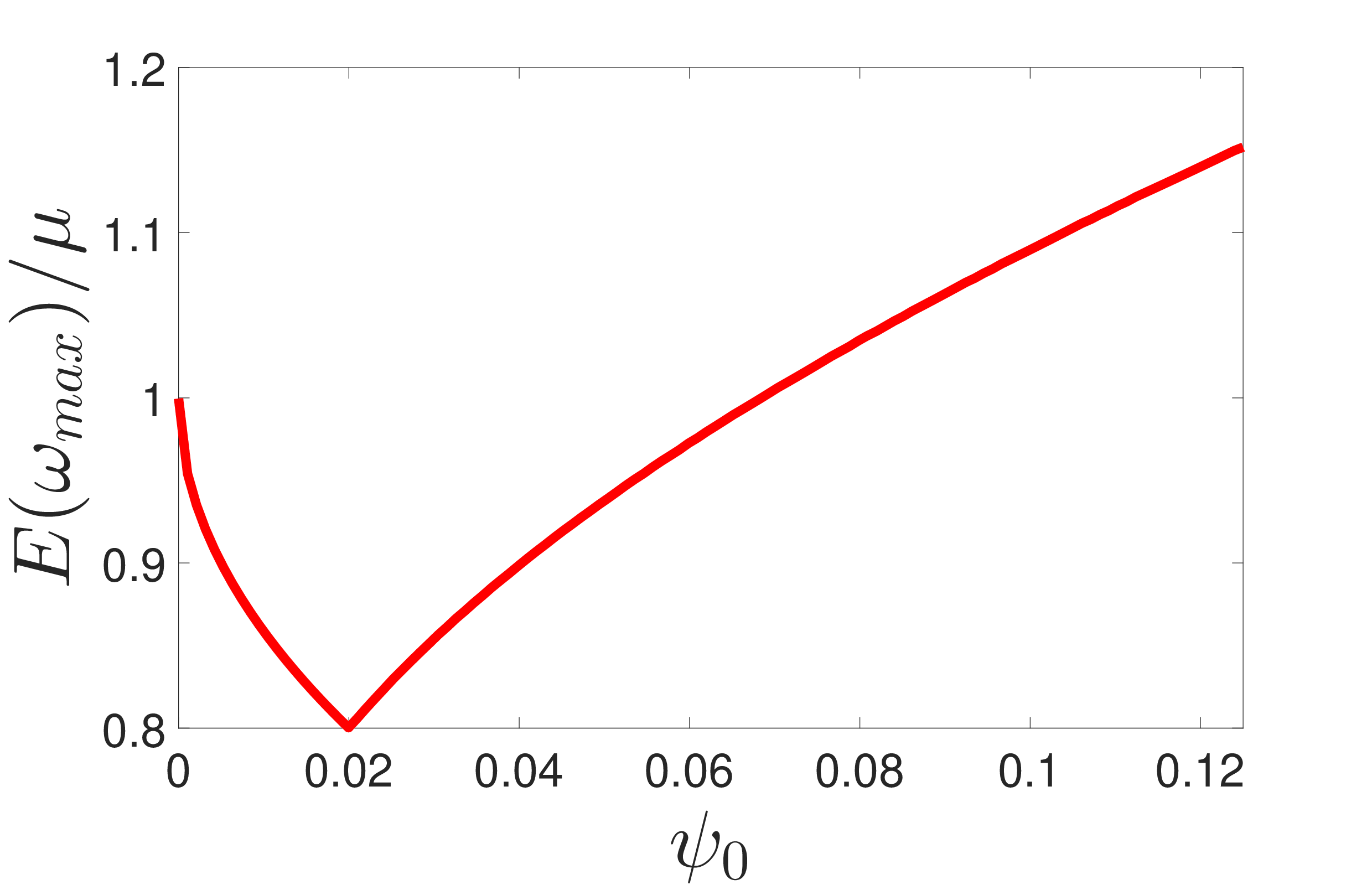}}
     \subfigure[]
          {\includegraphics[width=0.49\columnwidth]{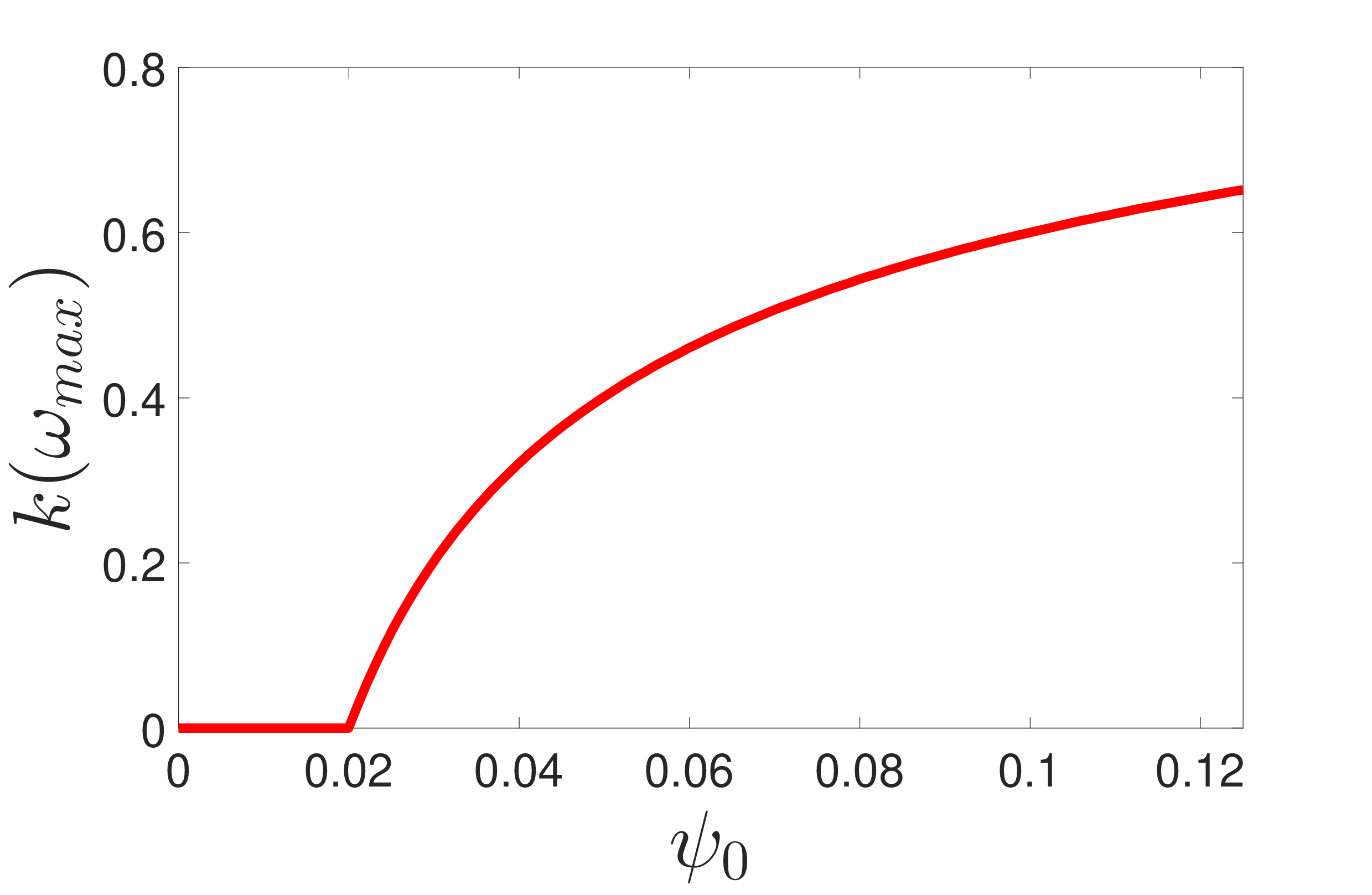}}
     \caption{(a) Bounce frequency $\omega_b$ dependence on energy $(k)$ and $\psi_0$. (b) Maximum bounce frequency as a function of $\psi_0$. (c), (d) Energy and $k$ corresponding to the maximum bounce frequency; for small $\psi_0$ the maximum frequency corresponds to the deeply trapped $k\simeq0$ particles.}
     \label{fig:ImageSize.testing.quarter}    
\end{center}
\end{figure*}

\begin{figure*}\centering
\begin{center}
     \subfigure[]
       {\includegraphics[width=0.49\columnwidth]{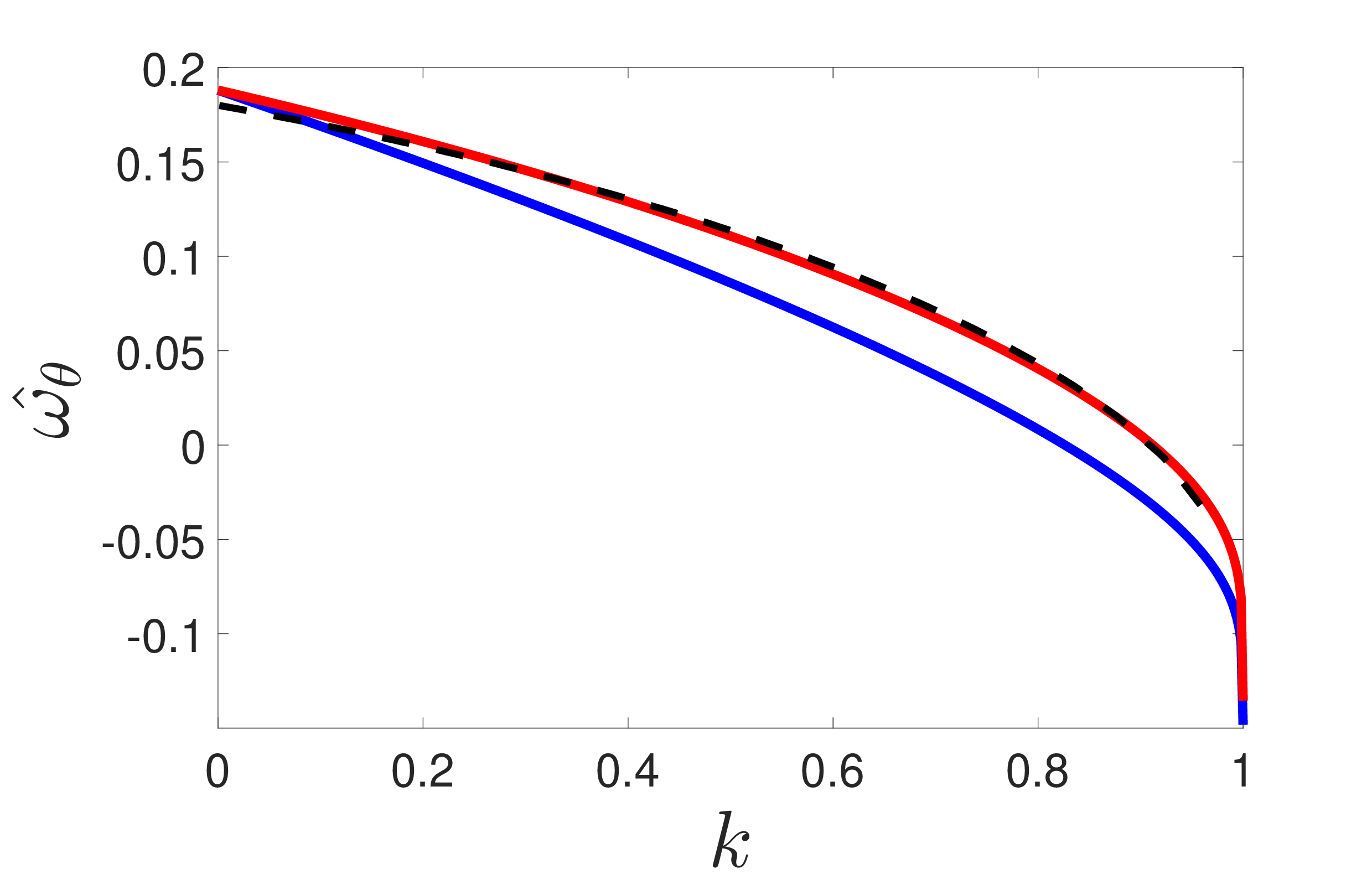}}
     \subfigure[]
          {\includegraphics[width=0.49\columnwidth]{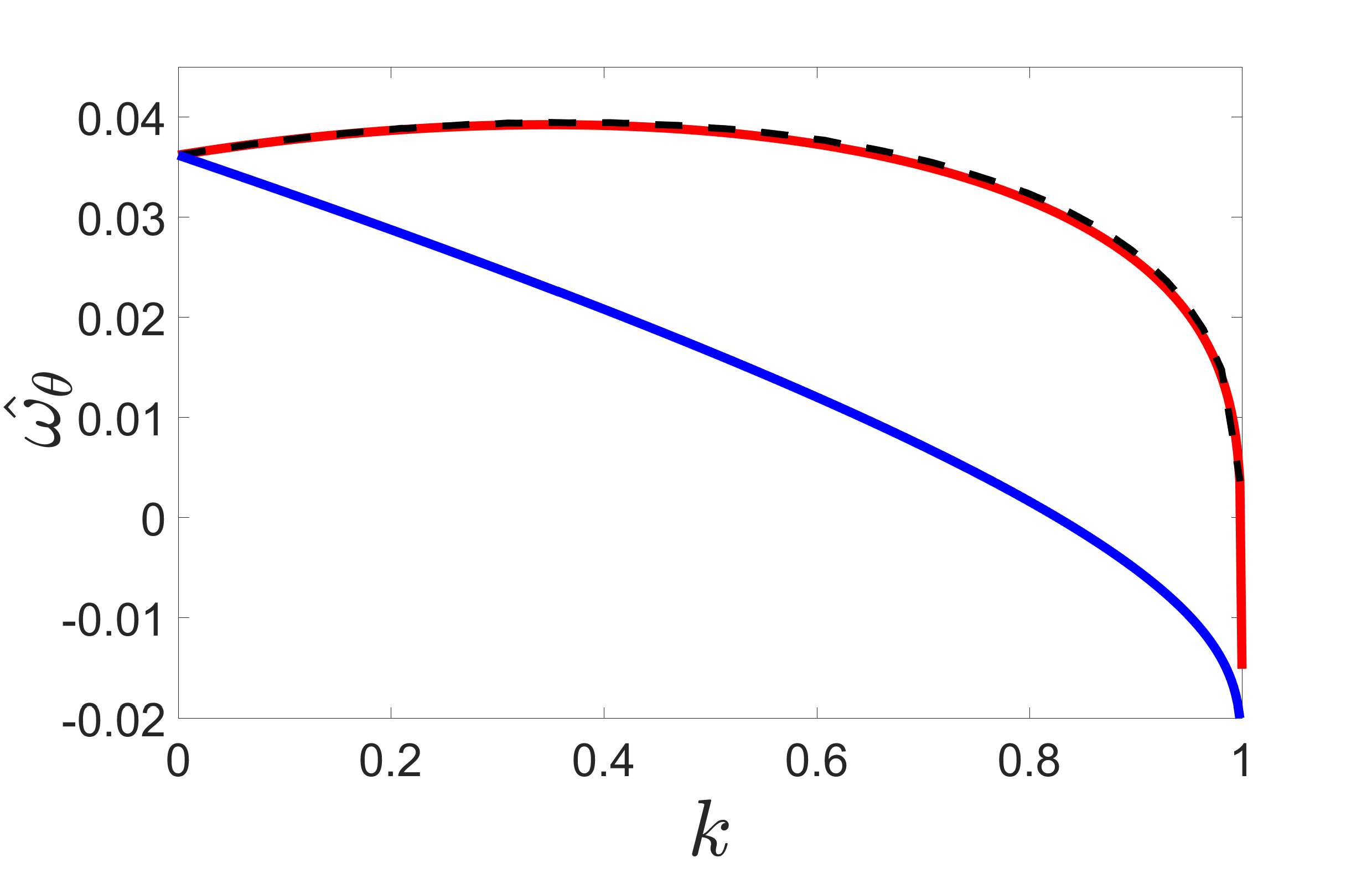}}
     \caption{Analytically (solid lines) and numerically (dashed lines) calculated bounce-averaged toroidal precession frequencies according to $H'_{ZDW}$ (blue line), $H_{ZDW}$ (red line) and $H$ [Eq. (\ref{H2})] (dashed line) for $q=1$, $\mu=10^{-4}$, and (a) $\psi_0=0.01$ $(\eta=-0.33)$, (b) $\psi_0=0.09$ $(\eta=-1.47)$.}
     \label{fig:ImageSize.testing.quarter}    
\end{center}
\end{figure*}

\begin{figure*}\centering
\begin{center}
     \subfigure[]
       {\includegraphics[width=0.49\columnwidth]{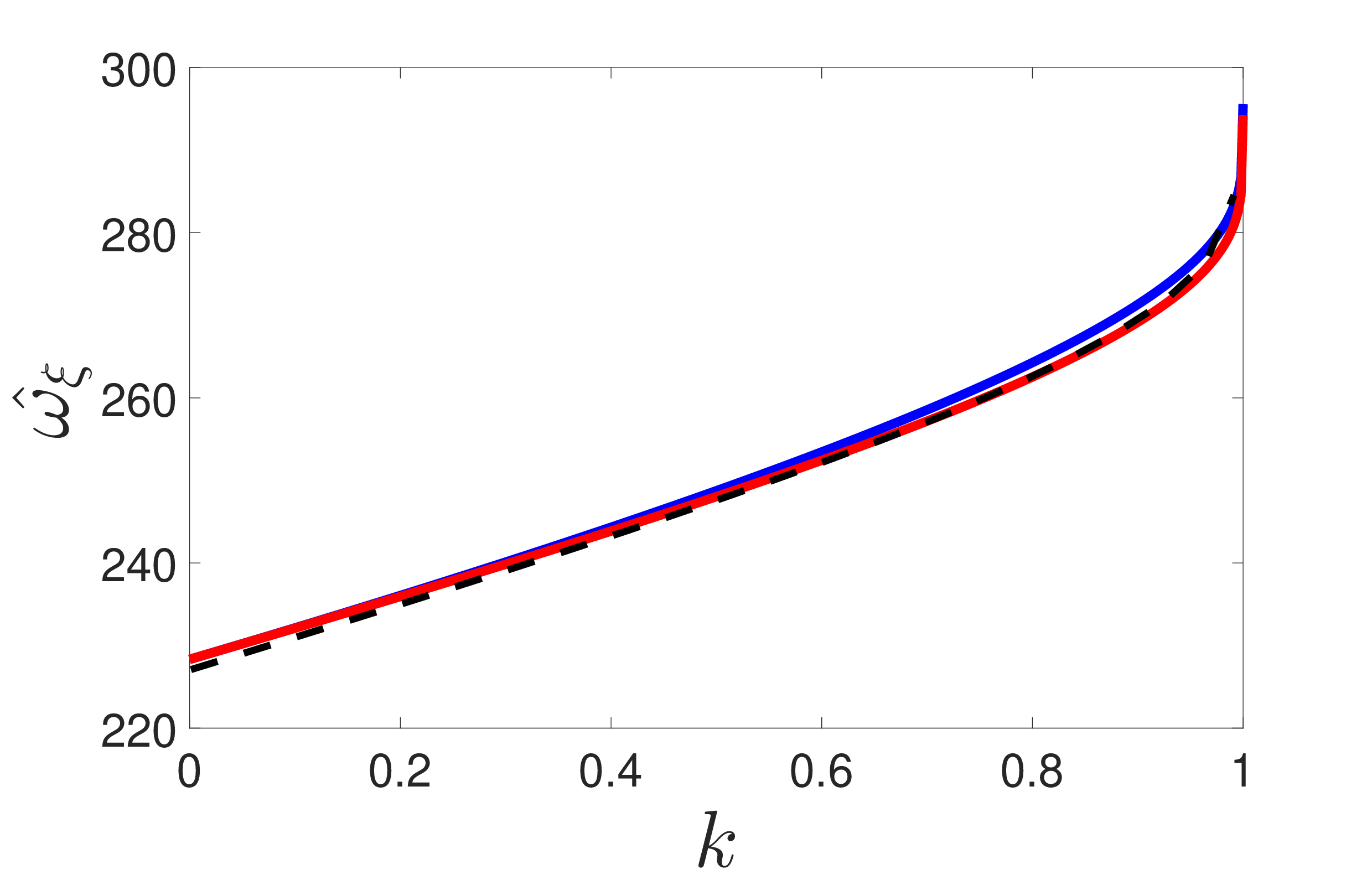}}
     \subfigure[]
          {\includegraphics[width=0.49\columnwidth]{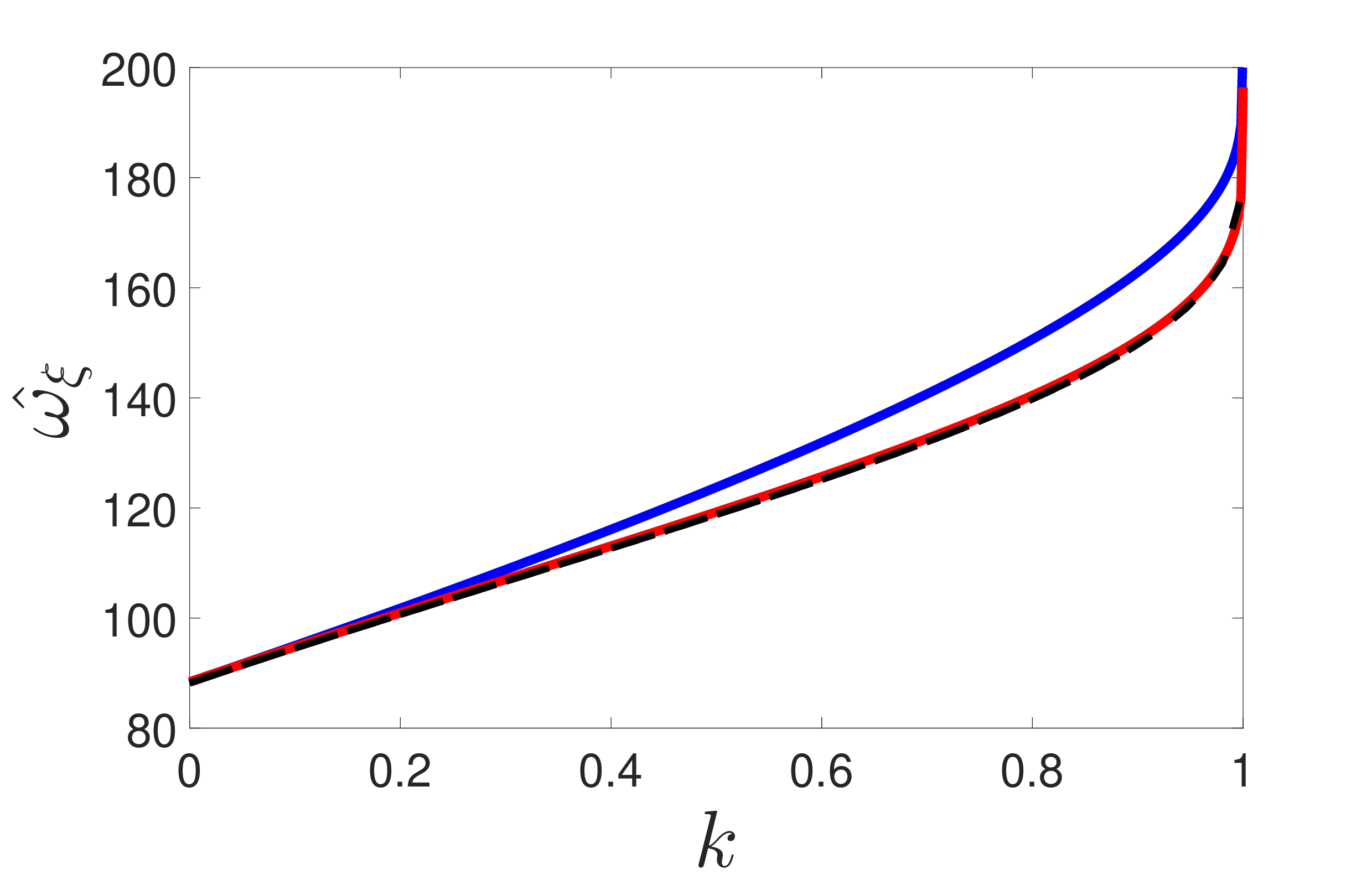}}
     \caption{Analytically (solid lines) and numerically (dashed lines) calculated bounce-averaged gyrofrequencies according to $H'_{ZDW}$ (blue line), $H_{ZDW}$ (red line) and $H$ [Eq. (\ref{H2})] (dashed line) for $q=1$, $\mu=10^{-4}$, and (a) $\psi_0=0.01$ $(\eta=-0.33)$, (b) $\psi_0=0.09$ $(\eta=-1.47)$.}
     \label{fig:ImageSize.testing.quarter}    
\end{center}
\end{figure*}

\end{document}